\documentclass[10pt,preprint]{aastex}
\shorttitle{M 2-9}
\shortauthors{Smith et al.}
\begin{document}

\title{Kinematic Structure of H$_2$ and [Fe~{\sc ii}] in the Bipolar
Planetary Nebula M 2-9}

\author{Nathan Smith\altaffilmark{1,2}}
\affil{Center for Astrophysics and Space Astronomy, University of
Colorado, 389 UCB, Boulder, CO 80309}

\author{Bruce Balick} 
\affil{Department of Astronomy, University of Washington, Box 351580,
Seattle, WA 98195}

%\and

\author{Robert D.\ Gehrz\altaffilmark{2}}
\affil{Astronomy Department, University of Minnesota, 116 Church St.\
SE, Minneapolis, MN 55455}

\altaffiltext{1}{Hubble Fellow; nathans@casa.colorado.edu}

\altaffiltext{2}{Visiting astronomer at the IRTF, operated by the
University of Hawaii under contract with NASA.}

\begin{abstract}

We present new high-dispersion long-slit infrared (IR) spectra of the
double-shell bipolar planetary nebula M~2-9 in the emission lines
[Fe~{\sc ii}] $\lambda$16435 and H$_2$ $v$=1--0 S(1) $\lambda$21218.
H$_2$ spectra reveal for the first time the kinematic structure of the
outer shell in M~2-9.  Kinematics of the inner shell, traced by
[Fe~{\sc ii}], resemble those of optical forbidden-lines like [N~{\sc
ii}] $\lambda$6583, although we note subtle differences.  [Fe~{\sc
ii}] and H$_2$ shells have expansion speeds roughly proportional to
distance from the star (``Hubble'' flows) and share the same dynamical
age of 1200--2000 yr, depending on the distance to M~2-9.  Thus, the
inner ionized lobes and outer molecular lobes, as well as the
molecular torus and ``outer loops'' measured by other observers, {\it
were all formed around the same time}.  Consequently, their nested
structure likely arises from an excitation gradient rather than
independent ejections.  H$_2$ and [Fe~{\sc ii}] emission is
distributed more uniformly than [N~{\sc ii}], and IR lines are not
dominated by the moving ionization pattern like visual-wavelength
lines.  We suggest that this is because IR lines of [Fe~{\sc ii}] and
H$_2$ are excited by relatively isotropic far-UV radiation (Balmer
continuum), whereas optical lines respond to a directed rotating beam
of Lyman continuum.  Finally, we highlight intriguing similarities
between M~2-9 and the Homunculus of $\eta$ Car, despite the different
central engines powering the two nebulae.

\end{abstract}

\keywords{circumstellar matter --- planetary nebulae: general ---
  planetary nebulae: individual: (M~2-9) --- stars: evolution ---
  stars: mass-loss}

\section{INTRODUCTION}

Among the many known examples of bipolar planetary nebulae (PNe;
Balick \& Frank 2002), the ``Butterfly'' nebula M~2-9 (Minkowski 1947)
is particularly striking.  Its smooth, almost cylindrical bipolar
lobes meet in a very tightly-pinched waist at the location of the
bright central star.  The remarkable optical morphology of M~2-9 has
been discussed by numerous authors (e.g., Allen \& Swings 1972;
Kohoutek \& Surdej 1980; Goodrich 1991; Schwarz et al.\ 1997; Balick
1999; Doyle et al.\ 2000).

M~2-9 may be a symbiotic PNe (Balick 1989) -- a member of a class of
tight-waisted bipolar nebulae with bright symbiotic stars or candidate
symbiotic stars as their central engines; other examples are Mz~3,
He~2-104, and Hubble~12 (e.g., Corradi et al.\ 2000; Schwarz et al.\
1989; Whitelock 1987).  M~2-9 and Mz~3 are near spectroscopic twins at
visual and infrared (IR) wavelengths (Swings \& Andrillat 1979; Hora
\& Latter 1994; Smith 2003), except that Mz~3 shows no IR emission
lines of molecular hydrogen (Smith 2003). M~2-9 is noteworthy in that
its bipolar lobes have almost perfectly smooth and thin boundaries in
{\it Hubble Space Telescope} ({\it HST}) images (Balick 1999), whereas
Mz~3 shows complex structure due to gas dynamic instabilities often
seen in PNe (Santander-Garcia et al.\ 2004; Guerero et al.\ 2004).

Even among this special class of PNe, M~2-9 is unique in exhibiting
rapid changes in the ionization structure of its bipolar lobes.  The
lobes are illuminated by a moving searchlight beam thought to be
caused by an asymmetrc rotating excitation source (either a collimated
jet or focused UV illumination) from a central binary star system.
This changes the emission-line appearance of the nebula over
timescales of years to decades (Doyle et al.\ 2000; Balick 1999;
Goodrich 1991; Kohoutek \& Surdej 1980; Allen \& Swings 1972).  The
nebula has several components: a bright nucleus containing the
putative symbiotic binary system (Lim \& Kwok 2000; Balick 1989,
1999), limb-brightened cylindrical bipolar lobes within
$\pm$25\arcsec\ of the nucleus where the unusual variability is seen,
and fainter polar loops extending $\pm$60\arcsec\ from the core (Solf
2000; Schwarz et al.\ 1997).  Bright condensations called ``ansae''
(Frank et al.\ 1996; Garcia-Segura 1997) are seen inside the polar
lobes as well, perhaps caused by shocks where fast jets strike the
interiors of the polar lobes (Balick 1999; Solf 2000).  The bright
bipolar lobes are dusty, giving rise to extended thermal-IR emission
(Smith \& Gehrz 2005) and polarization in scattered light (King et
al.\ 1981; Trammell et al.\ 1995).

Here we investigate kinematics of the brightest parts of the bipolar
lobes using IR spectroscopy.  Hora \& Latter (1994; HL94 hereafter)
noted a striking double-shell structure in M~2-9, with the smaller
pair of bipolar lobes emitting bright [Fe~{\sc ii}] emission nested
inside an outer pair of lobes comprising a thin skin of H$_2$ emission
(see also Kastner et al.\ 1996).  Aside from faint [O~{\sc i}]
emission, the outer lobes are seen only in scattered light at visual
wavelengths (Balick 1999).  This structure, with [Fe~{\sc ii}] bulbs
inside a thin H$_2$ skin, is similar to the double-shell structure in
the nebula surrounding $\eta$ Carinae (Smith 2002).  [Fe~{\sc ii}]
bulbs inside a larger H$_2$ nebula are also seen in other PNe, such as
Hubble~12 (Welch et al.\ 1999; Hora \& Latter 1996) and NGC~7027
(Latter et al.\ 2000).  In {\it HST} images, the outer and inner
shells of M~2-9 distinguish themselves in low- and high-excitation
emission lines like [O~{\sc i}] and [O~{\sc iii}], respectively
(Balick 1999).  Given their segregated geometry in images, we wish to
determine if the inner and outer lobes of M~2-9 have similar
kinematics; i.e. did they arise from separate and distinct ejections,
or were they formed in just a single event, so that their stratified
structure arises instead from density gradients and radiative transfer
effects like in the case of $\eta$~Car?\footnote{The case of
$\eta$~Car is even more complex, however.  Note that here we are
referring to the inner walls of the main Homunculus nebula seen in
[Fe~{\sc ii}] emission, which have the same age as the outer H$_2$
skin (Smith 2002).  We are not referring to [Fe~{\sc ii}] emission
from the ``Little Homunculus'', which was ejected in a separate event
50 years later (Smith 2005).}  In lieu of independent age
determinations from proper motion measurements of each shell component
in M~2-9, we investigate the kinematics of these two shells using
Doppler shifts in high-dispersion spectra of [Fe~{\sc ii}]
$\lambda$16435 tracing the inner shell, and H$_2$ $\lambda$21218
tracing the outer shell.

\section{OBSERVATIONS}

\subsection{IR Images}

Figure 1 shows a three-color composite image of M~2-9, included here
to illustrate the position of the spectroscopic aperture relative to
the H$_2$ and [Fe~{\sc ii}] emission.  H$_2$ $\lambda$21218 emission
is red, Br$\gamma$ and continuum is green, and [Fe~{\sc ii}]
$\lambda$16435 is blue.  Extended green emission is very faint in
Figure 1 because Br$\gamma$ emission seen there is mainly reflected
light (e.g., Trammell et al.\ 1995), whereas strong intrinsic [Fe~{\sc
ii}] and H$_2$ lines are emitted by the lobes.  These images were
taken on the University of Hawaii 2.2 m telescope on 1997 April 25
using the QUIRC instrument (Hodapp et al. 1996),\footnote{See also
{\url http://www.ifa.hawaii.edu/instrumentation/quirc/quirc.html}.}
and the data reduction was similar to that described by HL94.

\subsection{IR Spectroscopy}

On 2004 Aug 20 we observed M~2-9 with the high-resolution spectrograph
CSHELL (Greene et al.\ 1993) mounted on NASA's Infrared Telescope
Facility (IRTF).  CSHELL has a 256$\times$256 SBRC InSb array with a
spatial pixel scale of 0$\farcs$2. Only 160 pixels are illuminated in
the spatial direction, yielding an effective slit length of roughly
30\arcsec.  We used a slit width of 0$\farcs$5, providing a spectral
resolving power of $R\approx$43,000 or 7 km s$^{-1}$.  The circular
variable filter (CVF) wheel isolates a single order of the echellette,
and we chose central wavelengths corresponding to the bright [Fe~{\sc
ii}] $\lambda$16435 emission line (vacuum wavelength 16439.98 \AA) and
the $v=1-0$ S(1) line of molecular hydrogen at 21218.36 \AA\ (vacuum).

We oriented the slit at P.A.=0\arcdeg\ and positioned it roughly
0$\farcs$5 east the central star (see Fig.\ 1) in order to move the
peak emission from the central star out of the slit so that we could
take longer exposures of the faint nebula.  We used total on-source
integration times of 40 and 35 minutes for [Fe~{\sc ii}] and H$_2$,
respectively, and sky-subtraction was accomplished with identical
observations of an off-source position.  The resulting 2-D spectra are
shown in Figure 2.  Wavelengths were calibrated using telluric
absorption lines, adopting wavelengths in the telluric spectrum
available from NOAO. We also used OH airglow lines for wavelength
calibration in the 2~$\micron$ region (wavelengths kindly provided by
R. Joyce; private comm.).  Uncertainty in the absolute wavelength
calibration is roughly $\pm$1 km s$^{-1}$.  Velocities in Figure 2 are
heliocentric, and have been corrected for the Earth's motion (to
convert to LSR velocities, add 14.7 km s$^{-1}$).  We did not attempt
to flux calibrate the spectra due to intermittent clouds on the night
the data were taken.

In Figure 3 we compare our new CSHELL spectra of [Fe~{\sc ii}] and
H$_2$ with similar long-slit spectra of [N~{\sc ii}] $\lambda$6583,
obtained in November 1998 and described already by Balick
(1999). These observations were obtained at roughly the same slit
position as is shown in Figure 1.

\section{RESULTS}

\subsection{Kinematic Structure of H$_2$ and [Fe~{\sc ii}]}

One of the most striking properties of M~2-9 in images (Figure 1) is
the double-shell structure, with an inner shell of [Fe~{\sc ii}] and
an outer shell of H$_2$.  This structure is evident in kinematics seen
in long-slit spectra in Figure 2 as well.  Qualitatively, the similar
morphology seen in images and in long-slit spectra suggests that the
nebular expansion is homologous.\footnote{Our slit is not exactly
along the polar axis, but it is close enough that this holds true
within the resolution of our data.}  This is especially true for the
outer shell of H$_2$ emission. Overall, the [Fe~{\sc ii}] is confined
within the H$_2$ walls in long-slit spectra and appears to follow the
same trend of increasing expansion velocity with distance from the
star, so that both H$_2$ and [Fe~{\sc ii}] follow the same basic
``Hubble'' flow.  The similarity between morphology in H$_2$ images
and kinematic structure in spectra implies that second-order effects
like non-radial motion at shock interfaces do not dominate the overall
shaping of the outer nebula.  Homologous expansion like this would
usually indicate that all the material comprising the H$_2$ shell was
ejected at roughly the same time (see \S 3.5). [Fe~{\sc ii}] emission
shows the apparent Doppler shift increasing with separation from the
star; it does not resemble the body of a steady jet, where we would
expect to see constant or even decreasing Doppler shifts at increasing
distance from the star.

While the kinematic structure of H$_2$ emission suggests homologous
expansion, its brightness distribution is not perfectly symmetric.
The near sides of the N and S polar lobes are 3.3 and 3.0 times
brighter than their respective far sides (see Fig.\ 4).  While this
would normally be attributed to extinction of the far side of the
nebula by dust, we cannot justify this assumption in the case of M~2-9
because the optical emission is known to vary with time due to a
rotating illumination source, so the intrinsic brightness of the front
and back sides may be very different.  This is discussed further in \S
4.2.  Also, there are some spatial asymmetries in the H$_2$
brightness; for example, in the S polar lobe, the brightest H$_2$
emission on the far side is much farther from the star than on the
near side.

\subsection{Comparison with [N~{\sc ii}] Kinematics and the Emission Knots}

Since [Fe~{\sc ii}] emission comes from the same inner lobes seen in
{\it HST} images of optical lines like [N~{\sc ii}], it is instructive
to compare their kinematics.  Figure 3 shows similar long-slit spectra
for [N~{\sc ii}] $\lambda$6583 at the same spatial offset from the
star (Fig.\ 3$b$; see also Icke et al.\ 1989), as well as [N~{\sc ii}]
contours superposed on [Fe~{\sc ii}] (Fig.\ 3$a$) and H~$_2$ contours
over [N~{\sc ii}] emission (Fig.\ 3$c$).

In general, we see that [Fe~{\sc ii}] and [N~{\sc ii}] occupy the same
kinematic space inside of and sheathed by the H$_2$ emission.
However, Figure 3$a$ shows a subtle but important difference: {\it the
knots N2 and S2 are not seen in} [Fe~{\sc ii}].  Knots N2 and S2 are
the most prominent signposts of the moving excitation beams that make
M~2-9 so unique.  Their absence in [Fe~{\sc ii}] implies that this IR
line is not influenced as strongly by the asymmetric rotating
excitation source.  While there are some concerns due to temporal
effects\footnote{The [Fe~{\sc ii}] observation was obtained about six
years after the [N~{\sc ii}] observation, and the N2 knots have moved
somewhat during this time. However, based on their previous rate of
motion, we would still expect these features to be easily included
within the CSHELL aperture.  Obviously, however, new high-resolution
imags of [Fe~{\sc ii}] and continued monitoring of the optical lines
would be useful to justify these comments.}, the N2 and S2 knots are
missing in [Fe~{\sc ii}] images (Fig.\ 1), while they dominate the
appearance of [N~{\sc ii}] images.  Their absence in [Fe~{\sc ii}] may
help unravel the nature of the excitation source, as discussed in \S
4.

Knots N1 and S1 are faintly visible in [Fe~{\sc ii}] emission in
Figure 2$a$.  Both features are blueshifted and have no observable
redshifted counterparts, indicating that the N1 and S1 knots are
located only on the near side of the nebula.  Interestingly, Figures
2$c$ and 3$c$ show that the N1 and S1 knots in [Fe~{\sc ii}] and
[N~{\sc ii}] appear at the same velocity as H$_2$ and mark the spatial
position where the brightest H$_2$ emission abruptly ends on the near
side of the polar lobes.  Do the N1 and S1 knots mark a position in
the outer lobes where the H$_2$ is dissociated and the gas is
partially ionized?  {\it HST} images certainly do not give the
impression that the N1 and S1 knots are part of the outer shell
(Balick 1999), so this is an interesting mystery.  In any case, N1 and
S1 are probably not isolated knots, but rather the illuminated part of
a ring or inner edge of a cylinder.

While [Fe~{\sc ii}] and [N~{\sc ii}] emission from S1 line up
perfectly in Figure 3$a$, the two emission lines from N1 show a
peculiar velocity difference of almost 10 km s$^{-1}$.  On the other
hand, [Fe~{\sc ii}] emission from N3 overlaps perfectly with [N~{\sc
ii}], while S3 shows a spatial difference of almost 1\arcsec\ and a
velocity difference of a few km s$^{-1}$ compared to [N~{\sc ii}].  S3
is also relatively much fainter than N3 in [Fe~{\sc ii}].  These
peculiar differences may be an effect of subtle changes in slit
position or temporal effects (N1 and S1 also exhibit the consequences
of the rotating excitation source), but further study is needed.

\subsection{Emission from the Nucleus}

Figures 2$b$ and 4$a$ reveal H$_2$ $\lambda$21218 emission in the
nucleus of M~2-9, in the immediate vicinity of the central star
(within 1000 AU).  HL94 did not detect any H$_2$ emission from the
central star in lower resolution spectra --- our detection of H$_2$ in
this study is probably because the CSHELL slit aperture was offset
0$\farcs$5 east of the star, excluding much of the central star's
bright continuum emission, and including more extended emission.  Its
detection in an offset position suggests that the H$_2$ emission
arises from a region within at least $\sim$1\arcsec\ of the star
perpendicular to the polar axis.

The contours in Figure 2$c$ show that the nuclear H$_2$ emission is
also extended about $\pm$0$\farcs$5 along the slit, showing a ``tilt''
in its kinematic structure, with blueshifted emission toward the north
and redshifted emission toward the south.  This indicates that the
H$_2$ emission arises in an equatorial molecular disk or torus, rather
than in polar material (polar material is redshifted toward the
north).  This is consistent with the detection of a CO torus within
$\pm$3\arcsec\ of the star by Zweigle et al.\ (1997). The H$_2$
emission arises closer to the central star, perhaps marking the
irradiated inner edge of this larger torus.

The extracted spectrum in Figure 4$a$ (which includes all emission
within $\pm$1$\farcs$5) reveals a narrow but resolved H$_2$ emission
line near the systemic velocity of M~2-9, with a FWHM of $\sim$25 km
s$^{-1}$.  The width of this line is consistent with the conjecture
that it is associated with the same molecular torus observed by
Zweigle et al.\ (1997).  This core region also shows strong [Fe~{\sc
ii}] emission line that is broader and more asymmetric than the H$_2$
emission line.  The asymmetric profile of [Fe~{\sc ii}] is similar to
that of [N~{\sc ii}] $\lambda$6583 in the central star (Solf
2000). The core [Fe~{\sc ii}] emission is not spatially resolved in
our data.

We see two high-velocity components of H$_2$ emission, about 5 times
fainter than the centroid component, at roughly $\pm$55 km s$^{-1}$
with respect to the systemic velocity.  Like the centroid component of
H$_2$, these two high velocity components are also spatially resolved
(see the contours in Figs.\ 2$c$ and 4$c$), and follow the same
kinematic trend as the tilt of the centroid component, indicating that
they are equatorial.  The faint blueshifted peak at +10.9 km s$^{-1}$
(heliocentric) is offset $\sim$0$\farcs$4 north, and the redshifted
component at +123.7 km s$^{-1}$ is offset 0$\farcs$3 south.  Solf
(2000) detected analogous high velocity components in optical emission
lines that were spatially extended by an amount similar to the fast
components of H$_2$ that we observe. However, the spatially-resolved
kinematic structure of those atomic lines showed the opposite
kinematic trend -- i.e. they were redshifted to the north and
blueshifted to the south, consistent with a fast {\it polar} outflow
instead of an equatorial outflow.  Solf (2000) proposed that the
high-velocity components in atomic lines reveal a bipolar microjet
near the star.  The faintest broad [N~{\sc ii}] emission within
2\arcsec\ of the star in Figure 3$b$ (blueshifted to the south and
redshifted to the north) extrapolates back to the position of the star
at roughly the same velocity as the fast H$_2$ emission at $\pm$55 km
s$^{-1}$.  Balick (1999) pointed out that if this material arises in a
bipolar jet near the star, then the jet decelerates rapidly even
though no strong radiative shock emission is seen.  One alternative
interpretation is that the [N~{\sc ii}] emission originates in fast
material in the walls of the polar lobes, and the H$_2$ emission
arises in a disk or ring where the [N~{\sc ii}] emission meets the
equator; in that case, the rapid deceleration seen in [N~{\sc ii}] is
an effect of the orientation angle rather than deceleration along the
same path.  The difference in H$_2$ velocities between the slow and
fast component at the same position is puzzling, since the implied
shock velocity is strong enough to dissociate H$_2$.  Then again, the
high velocity H$_2$ emission we observe may be something more
complicated like the equatorial H$_2$ emission in the Egg nebula (Cox
et al.\ 2000; Sahai et al.\ 1998; HL94).  In any case, one might infer
that future observations of H$_2$ and atomic lines with high spatial
resolution techniques would have the potential to reveal the wind
collimation mechanism near the central star.

\subsection{The Systemic Velocity of M~2-9}

We find three independent potential indicators of M~2-9's systemic
radial velocity.  These are: 1) the average of the velocities for
[Fe~{\sc ii}] emission in the N3 and S3 knots at the ends of the polar
lobes, 2) the average of the centers of expansion (i.e. halfway
between the blue and redshifted surfaces) for the two polar lobes seen
in narrow H$_2$ emission at similar spatial offsets from the equator,
and 3) the centroid velocity of the H$_2$ emission from the central
star, presumably from a circumstellar molecular disk in the core.  Of
these three, the second is arguably the most reliable (as is the case
for $\eta$~Car; Smith 2004), given the narrow emission components and
the apparent symmetry of the nebula in H$_2$ images (HL94). The
relevant velocity measurements are collected in Table 1, where the
three methods just mentioned are shown in bold font.  In Table 1, each
measurement is the average of a Gaussian fit and a flux-weighted
centroid; the difference between these two measurement methods was
always less than $\pm$0.5 km s$^{-1}$, which is less than the
uncertainty in the wavelength calibration of $\pm$1 km s$^{-1}$.

Taking the average of the three bold estimates in Table 1 gives
V$_{sys}$=+69.2 km s$^{-1}$ for the heliocentric systemic velocity of
M~2-9.  Measured with respect to the local standard of rest, the
systemic velocity would be V$_{LSR}$=+83.9 km s$^{-1}$.  This agrees
with measurements of the molecular torus around M~2-9, which imply a
systemic velocity at LSR values of +80 to +81 km s$^{-1}$ (Bachiller
et al.\ 1990; Zweigle et al.\ 1997).

\subsection{The Dynamical Age of M~2-9}

Figure 2$b$ gives the first reliable information about the kinematics
of the outer molecular shell in the bipolar lobes of M~2-9; comparing
its expansion speed to its current size gives valuable clues to the
age of this component of the nebula.  Because of the degeneracy
between the uncertain distance to M~2-9 and the lack of proper motion
measurements for the outer shell, we cannot give an accurate value for
the {\it absolute} age of the outer shell from our data.  However, we
can constrain the {\it relative} dynamical age of the molecular shell
compared to the ionized gas in the inner lobes of M~2-9 for an assumed
distance.  Given the double-shell structure of M~2-9 (H$_2$ vs.\
[Fe~{\sc ii}]), it would be very interesting to know if both
components were ejected at the same time (as is the case for $\eta$
Car; Smith 2002), or if the ionized gas is expanding into and
interacting hydrodynamically with an older and cooler molecular shell.
Solf (2000) proposed that the cool outer shell is three times older
than the hot inner shell, but that conjecture was based on velocities
in stellar-wind lines reflected by dust in the cool shell, rather than
a direct measure of the expansion of the cool shell itself.

The lateral expansion speed in the middle of the polar lobes
(difference between the blue and redshifted H$_2$ components in Table
1 and Fig.\ 4) is roughly 30 km s$^{-1}$ when corrected for the
73\arcdeg\ \ inclination angle of the nebula (e.g., Zweigle et al.\
1997).  In images (see Fig.\ 1 and HL94), the lateral size of the
outer H$_2$ shell (in the E/W direction) is 10$\farcs$5 for the N lobe
and 11$\farcs$5 for the S lobe.  We presume that the outer shell is
expanding homologously and with approximate axial symmetry.  Thus, the
average dynamical age of the H$_2$ shell would be 1750$\times$D$_{\rm
kpc}$ yr (with an uncertainty of less than 10\% beyond the uncertainty
of the assumed distance).  Thus, if D=650 pc (Schwarz et al.\ 1997),
then the age of the outer H$_2$ lobes is roughly 1140 yr.

Using optical forbidden lines, Solf (2000) found a dynamical age of
$\sim$1300 yr for the ionized inner shell adopting the same distance
of 650 pc to M~2-9.  This is comparable to the dynamical age of 1365
yr derived by Zweigle et al.\ (1997) for the molecular torus (Zweigle
et al.\ actually published an age of 2100 yr, but that was at a
distance of 1 kpc).  Both Schwarz et al.\ (1997) and Solf (2000)
derived an age of 1200-1300 yr for the faint outer loops 1\arcmin\
from the star along the polar axis.

The uncertainties in these various measurements are large enough to
make it likely that {\it all three components of the nebula are
coeval}, neglecting any potential acceleration or deceleration.  Based
on the nebula's present kinematics, our measurements make it seem
improbable that the outer molecular shell is significantly older than
the inner ionized gas in the polar lobes.  If the molecular shell was
3 times older than the inner ionized shell as Solf (2000) suggested,
then the observed velocities of the [Fe~{\sc ii}] emission in Figure 2
would not be confined within the walls of the H$_2$ emission.  {\it
Caveat}: Of course, the dynamical age we measure today is not
necessarily a meaningful indicator of the true ejection age.  For
example, the outer H$_2$ shell may have been ejected earlier than the
[Fe~{\sc ii}] bulbs if the [Fe~{\sc ii}] bulbs were ejected more
recently with higher speed, but have been decelerated upon interaction
with the more massive H$_2$ shell.

However, there is one nagging departure from this single-ejection
hypothesis.  The N3 and S3 knots seen in [Fe~{\sc ii}] appear to be
slightly faster than other material in the inner ionized shell,
suggesting one of two possible scenarios.  Either they are younger
than the other material emitting [Fe~{\sc ii}] because they have
higher speeds at the same distance from the star, or their
trajectories are tilted with respect to the assumed polar axis of
17$\arcdeg$.  We cannot rule out the second possibility, but N3 and S3
appear exactly along a projection of the polar axis in images (Balick
1999; HL94), so the younger age seems to be a more likely hypothesis.
The N3 and S3 knots have Doppler shifts of +15.2 and $-$13.0 km
s$^{-1}$ with respect to the systemic velocity (see Table 1), and they
are located at positional offsets of +15$\farcs$0 and $-$13$\farcs$9
from the central star, respectively.  If the inclination is
73$\arcdeg$ (Zweigle et al.\ 1997), the dynamical ages for the N3 and
S3 knots are roughly 1500$\times$D$_{\rm kpc}$ yr and
1600$\times$D$_{\rm kpc}$ yr, respectively. At 650 pc, their ages are
975 and 1040 yr -- about 15\% younger than the rest of the nebula.
With trajectories near 73\arcdeg\ from the line of sight, their
de-projected velocities are of order 50 km s$^{-1}$, yet N3 and S3
show no detectable proper motions in {\it HST}/WFPC2 images (Balick et
al.\ in prep.).  This is puzzling, and sugggests that their locations
may mark the position of a standing shock.

\section{DISCUSSION}

\subsection{Double Shells: Multiple Ejections or Excitation Structure?}

Our measurement that the inner [Fe~{\sc ii}] shell and the outer H$_2$
shell have roughly the same dynamical age argues against the simplest
explanation for the observed double-shell structure --- i.e., that
each shell is the product of a separate ejection event.  As noted
above, however, there is still room for multiple-ejections or
post-ejection shaping if the inner shell has been decelerated or the
outer shell accelerated by the interaction of the two shells, so that
they mimic a single ejection.  In either case, however, the
double-shell structure is still likely to be the result of different
layers in a photodissociation region through a single thick shell.

Naively, we might {\it expect} the stratified double-shell structure
seen in M~2-9 from pure photoexcitation by a central source, because
Fe$^+$ and H$_2$ should not occupy the same volume.  Fe$^0$ is ionized
to Fe$^+$ at $\sim$3.4 eV higher than the H$_2$ molecule is
dissociated.  This type of stratified structure might be seen, for
example, in the case of a strong source of far-UV photons incident
upon a thick, uniform-density shell, resulting in an ionization
gradient.  In their spectral analysis of multiple near-IR H$_2$ lines,
HL94 concluded that the near-IR molecular hydrogen emission in the
outer shell of M~2-9 is excited through absorption by the Lyman and
Werner bands in the far-UV.  Understanding the excitation of the
[Fe~{\sc ii}] lines is a more difficult matter, because shock
excitation, photoelectric heating through dust grains, or UV pumping
can produce a similar near-IR [Fe~{\sc ii}] spectrum.  Thus, to
determine if shocks excite the bright [Fe~{\sc ii}] emission from the
inner shell, more detailed shock models or photoionization
calculations are in order.  In any case, excitation in the far-UV must
play an important role for the [Fe~{\sc ii}] lines, since UV radiation
dominates the excitation of H$_2$ at even larger distances from the
star.

Even if the double-shell structure really represents two separate and
distinct shells inside one another, the existence of nested bipolar
lobes would not be unique in PNe.  The most relevant comparison,
perhaps, is the pair of bipolar nebulae at different radii seen around
He 2-104, both having the same dynamical age (Corradi et al.\ 2001).
It is not easy to imagine how nested coeval structures like these are
produced, and it is not clear that invoking a close mass-transfer
binary system will help (Balick \& Frank 2002).

\subsection{The Rotating Excitation Source}

Figure 1 implies that H$_2$ and [Fe~{\sc ii}] are more uniformly
distributed than familiar optical lines like H$\alpha$.  This
conjecture is supported by our spectra, where we detect emission from
the front and back of the nebula in H$_2$, and in [Fe~{\sc ii}] we do
not detect the strong front-back asymmetry evidenced by the N2 and S2
knots in [N~{\sc ii}].  While the H$_2$ emission is brighter on the
near side than on the back side of the nebula, this could be due in
part to dust extinction of the far side.  Also, it is clear from
images that H$_2$ shows a morphology very different from the
highly-directed moving pattern seen in H$\alpha$.  In any case, it is
fair to say that the IR lines of [Fe~{\sc ii}] and H$_2$ are not
influenced by the moving ionization source {\it to the same degree} as
lines like [O~{\sc iii}], H$\alpha$, and [N~{\sc ii}], where the
observed morphology is dominated by this phenomenon.  [Fe~{\sc ii}]
and H$_2$ (and [O~{\sc i}] at visual wavelengths) emission is more
uniform, whereas [O~{\sc iii}], H$\alpha$, and [N~{\sc ii}] are very
azimuth-dependent.  {\it What causes this difference?}

Suppose that the central engine of M~2-9 is a symbiotic binary system
composed of a very hot source and a cooler star with a dense mass-loss
wind.  By ``hot'' we mean that the star is the dominant source of
Lyman continuum photons in the system, and by ``dense'' we mean that
the wind is optically thick in the Lyman continuum.  If the powerful
wind of the cooler star overwhelms the wind of the hot companion, or
if a jet from the compact source clears a path through the dense wind
(e.g., Livio \& Soker 2001; Garcia-Arredondo \& Frank 2004), then
ionizing radiation from the hot star may be confined to escape only
through a narrow range of azimuthal angles (see Fig.\ 5).  This
direction of escape will rotate with time as the orbit proceeds,
sending a rotating searchlight beam of ionizing radiation out into the
nebula. This is essentially the scenario favored by Livio \& Soker
(2001).  However, the cool dense wind should be more transparent below
13.6 eV, so escape of Balmer continuum photons will be thwarted to a
lesser degree than the Lyman continuum.  Consequently, the Balmer
continuum radiation may escape from a larger azimuthal angle,
exhibiting only mild asymmetry by comparison.

Continuing with this qualitative scenario, the Lyman continuum photons
from the hot star that are intercepted by the
cool star's optically-thick wind will ionize some fraction of that
wind, regardless of the spectral type of the cool star.  A dense
partially-ionized wind would give rise to a strong P Cygni absorption
feature and prominent electron scattering wings.  In essence, the net
spectrum of the system would mimic that of a hot evolved star with a
strong wind.  Both of these are observed in the H$\alpha$ profile of
the central star (Balick 1989).  Electron scattering wings in a dense
wind extend to apparent Doppler shifts larger than actual gas
motions, so this would explain the very broad wings of H$\alpha$,
which extend several thousand km s$^{-1}$ to the blue and red.

In summary, then, the basic picture we favor is one where the infrared
[Fe~{\sc ii}] and H$_2$ emission is excited by relatively isotropic
far-UV (Balmer continuum) photons, and H$\alpha$ and other asymmtric
visual-wavelength lines result from a highly direction-dependent
rotating beam of Lyman continuum radiation from the central binary
system (Fig.\ 5).  The Balmer continuum need not be perfectly
isotropic to explain the observations, so long as it is more isotropic
than the ionizing radiation.  Without a strong source of Lyman
continuum, we may lack an explanation for the broad wings of H$\alpha$
from the central star.  A rotating UV illumination source also gives a
more straightforward explanation for the observed mirror symmetry in
the nebula above and below the equator, as opposed to the
point-symmetric structure expected from a precessing jet (Cliffe et
al.\ 1995; Garcia-Segura 1997; Livio \& Soker 2001; Garcia-Arredondo
\& Frank 2004).

\subsection{Comparing M~2-9 and Eta Carinae}

At first glance, comparing M~2-9 with $\eta$~Carinae might seem like
an exercise in futility.  After all, $\eta$ Car is more than two
orders of magnitude more luminous and almost two orders of magnitude
more massive than the central star(s) of M~2-9, and the Homunculus
nebula is about an order of magnitude more massive than the polar
lobes of M~2-9 (Smith et al.\ 2003; Smith \& Gehrz 2005).  $\eta$ Car
is an unstable hot supergiant teetering near the Eddington limit, it
has never been and will never be a cool asymptotic giant branch star,
and it will most likely end its life as an energetic supernova or
hypernova instead of as a PN.  These extreme differences in the
central stars make the similarities in their two circumstellar nebulae
even more remarkable.

{\it Double Shells}: As noted earlier, both M~2-9 and $\eta$~Car show
the same double-shell structure, with a thin outer molecular skin and
inner lobes seen in [Fe~{\sc ii}] (\S 3.1; HL94; Smith 2002).  This
points to similar excitation conditions and a similar density
structure in the hollow bipolar lobes of each nebula, where far-UV
from the central engine is attenuated by dust, allowing H$_2$ to form
in the outer lobes.  This scenario -- where Lyman continuum is quashed
by a dense stellar wind while Balmer continuum photons still escape --
certainly applies to $\eta$~Car.  Given this similarily, it is
peculiar that the clumpy mottled structure of the bipolar lobes around
$\eta$ Car is so different from the smooth structure in M~2-9.

{\it Hubble Flows}: In both objects, the kinematics seen in long-slit
spectra resemble the morphology in images, which is equivalent to
saying that they exhibit homologous expansion or ``Hubble'' flows.
This suggests either that most of the mass seen in images was ejected
in a single outburst (ejected on a timescale small compared to the age
of the nebula), or that material ejected in more than one event has
interacted dynamically to mimic a coeval outflow.  We know that the
former is true for $\eta$~Car, because proper motions and linear
expansion give an ejection date during the visually-observed outburst
in the 19th century (Smith \& Gehrz 1998).  In any case, the Hubble
flow observed in M~2-9 classifies it with several other bipolar
nebulae exhibiting Hubble flows (e.g., Corradi 2004; Balick \& Frank
2002; Corradi \& Schwarz 1993).

{\it Pinched Waists and Dust Tori}: Both nebulae have tightly-pinched
waists with the polar lobes meeting at a bright, obscured central
star.  In the case of $\eta$~Car, the polar lobes do not converge all
the way to the star, but instead meet at the equator to form a hot
dusty ring or torus with a radius of a few 10$^3$ AU (Smith et al.\
2003).  In M~2-9, the outer lobes also meet at the equator to form a
molecular ring (Zweigle et al.\ 1997), and M~2-9 appears to harbor a
warm dusty circumstellar torus (Smith \& Gehrz 2005).

{\it Rotating Excitation Source}: M~2-9 is unique compared to other
PNe -- even bipolar and symbiotic PNe with a similar tight-waisted
morphology.  The ionzation structure in its polar lobes changes on
short timescales (decades), where the brightest emission from ionized
gas appears to move laterally, in a direction perpendicular to the
polar axis (Doyle et al.\ 2000; Balick 1999; Goodrich 1991; Kohoutek
\& Surdej 1980; Allen \& Swings 1972).  This seems to be caused by a
source of ionization escaping the nucleus in a preferred direction
that rotates with time, as described in the previous section.
Analogous behavior has been seen recently in $\eta$~Carinae (Smith et
al.\ 2004), although its variable ``Purple Haze'' is revealed as
scattered UV continuum light rather than changing tracers of ionized
gas like the [O~{\sc iii}] emission in M~2-9.  In both objects, the
rotating excitation suggests that the central engine is a binary
system, and that the stars are close enough that the dense wind of one
star (an asymptotic giant branch star in M~2-9, or the luminous blue
variable in $\eta$ Car) confines the UV radiation or wind from the
secondary to escape only in certain directions as the stars orbit
around one another.

\acknowledgments \scriptsize

We thank Joe Hora and Bill Latter for providing the [Fe~{\sc ii}],
Br$\gamma$, and H$_2$ images used to make Figure 1, as well as helpful
comments on the manuscript.  N.S.\ was supported by NASA through grant
HF-01166.01A from the Space Telescope Science Institute, which is
operated by the Association of Universities for Research in Astronomy,
Inc., under NASA contract NAS5-26555.  R.D.G.\ was supported by NASA,
the NSF, the United States Air Force, and the Graduate School of the
University of Minnesota.

%
% REFERENCES

% TABLE 1 --- 
\begin{deluxetable}{llc}
%\tabletypesize{\normalsize}
\tabletypesize{\scriptsize}
\tablecaption{Heliocentric velocities in M 2-9\tablenotemark{a}}
\tablewidth{0pt}
\tablehead{
 \colhead{Line} &\colhead{Feature}  &\colhead{V (km s$^{-1}$)} 
}
\startdata

$[$Fe~{\sc ii}]	&N3 	 	&+84.4	\\
$[$Fe~{\sc ii}]	&S3 	 	&+56.2	\\
$[$Fe~{\sc ii}]	&N3+S3 average 	&{\bf +70.3}	\\

H$_2$	 	&star (blue) 	&+10.9	\\
H$_2$	 	&star (center) 	&{\bf +68.2}	\\
H$_2$	 	&star (red) 	&+123.7	\\
H$_2$	 	&N lobe (blue) 	&+57.5	\\
H$_2$	 	&N lobe (red) 	&+89.2	\\
H$_2$	 	&N lobe (avg.) 	&+73.4	\\
H$_2$	 	&S lobe (blue) 	&+49.5	\\
H$_2$	 	&S lobe (red) 	&+80.2	\\
H$_2$	 	&S lobe (avg.) 	&+64.8	\\
H$_2$	 	&N+S average 	&{\bf +69.1}	\\

\nodata	 	&V$_{sys}$	&+69.2	\\

\enddata
\tablenotetext{a}{Uncertainty in measured velocities is dominated by
$\pm$1 km s$^{-1}$ uncertainty in the original wavelength
calibration. Numbers in bold represent likely estimates for the
systemic velocity; the average of these three is given at the bottom
of the table as V$_{sys}$.}
\end{deluxetable}

% FIGURE 1 ---------- 
\begin{figure}
\epsscale{0.5}
\plotone{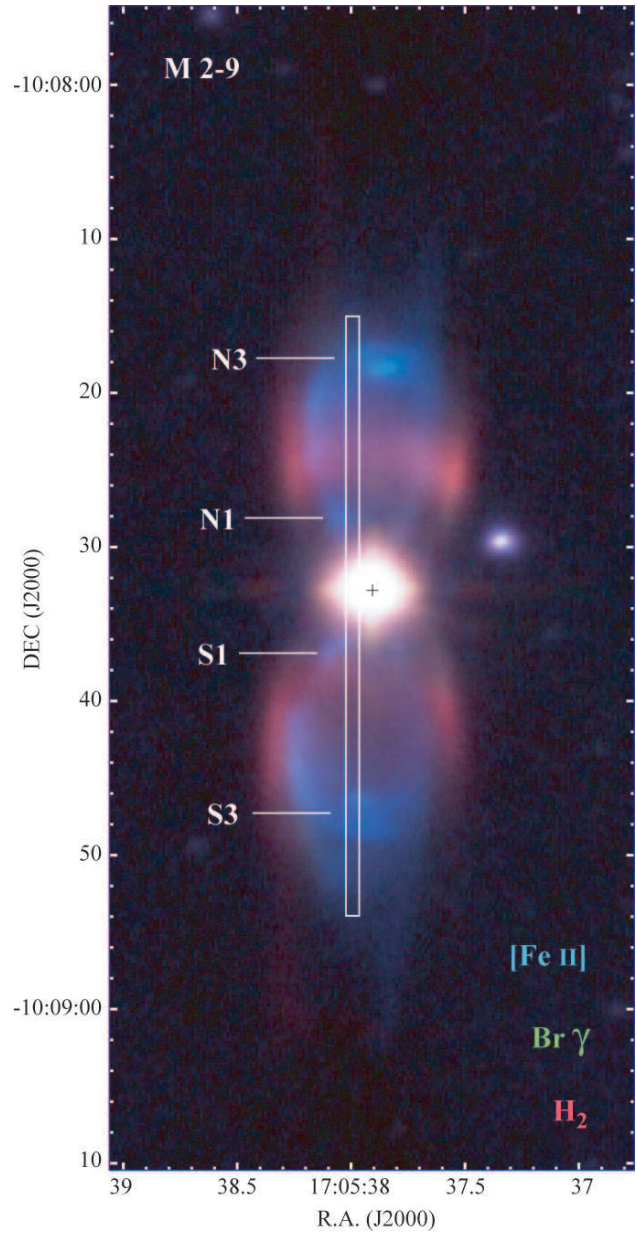}
\caption{Three-color image of M~2-9, with H$_2$ $\lambda$21218
  emission in red, Br$\gamma$ and continuum in green, and [Fe~{\sc
  ii}] $\lambda$16435 in blue.  The position of the CSHELL slit
  aperture is indicated.}
\end{figure}

% FIGURE 2 -----------
\begin{figure}
\epsscale{0.8}
\plotone{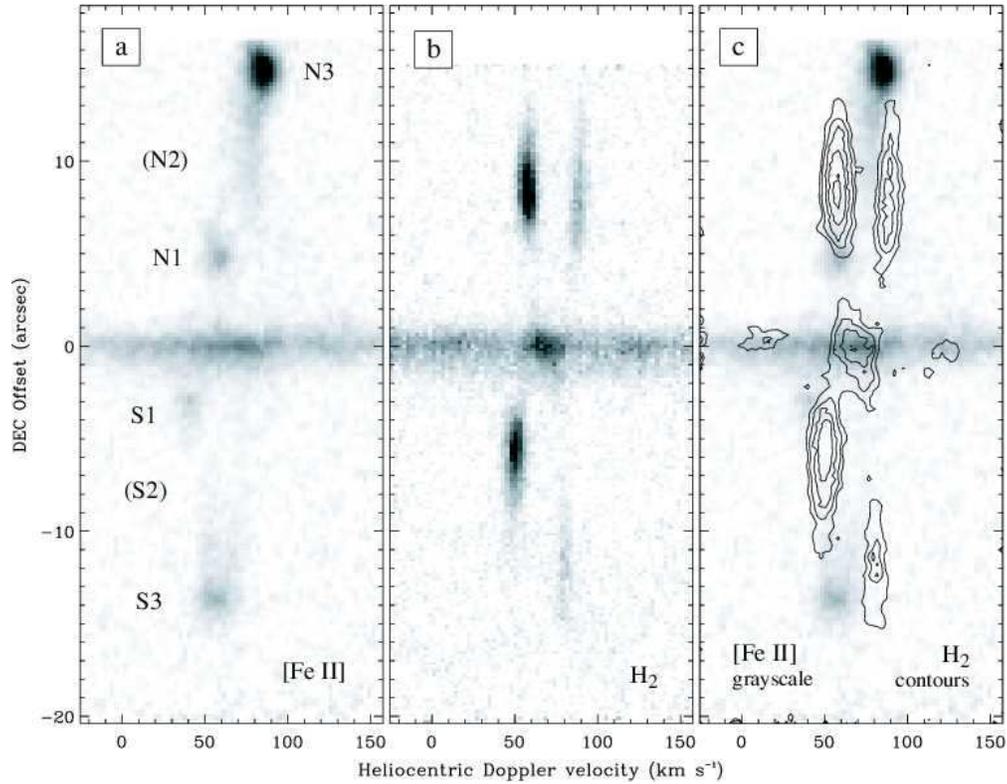}
\caption{Long-slit spectrograms of [Fe~{\sc ii}] and H$_2$, oriented
  roughly along the polar axis of M~2-9.  Panel (a) shows [Fe~{\sc
  ii}] $\lambda$16435 emission, Panel (b) shows H$_2$ $v$=1-0 S(1)
  $\lambda$21218, and Panel (c) shows H$_2$ contours over the [Fe~{\sc
  ii}] emission in grayscale.  In each panel, the scattered continuum
  light along the full length of the slit has been suppressed to
  enhance the extended emission-line structure.}
\end{figure}

% FIGURE 3 -----------
\begin{figure}
\epsscale{0.8}
\plotone{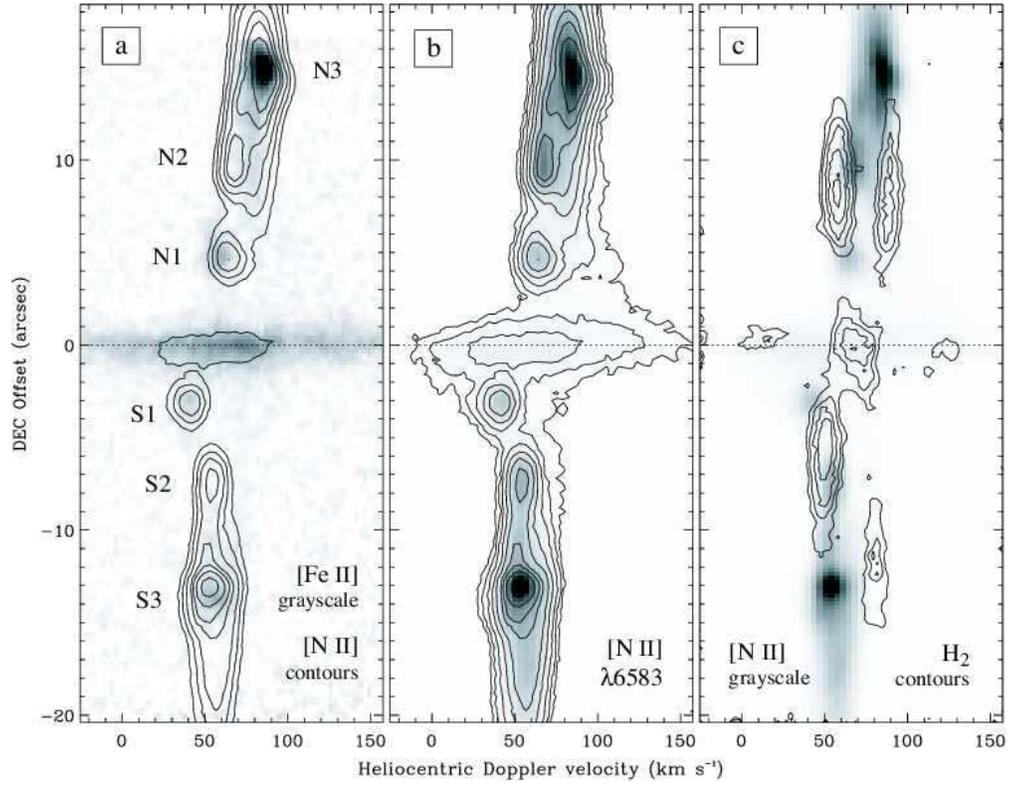}
\caption{Same as Figure 2, but comparing the infrared [Fe~{\sc ii}]
  and H$_2$ emission with optical [N~{\sc ii}] $\lambda$6583 emission.
  Panel (a) shows [Fe~{\sc ii}] $\lambda$16435 emission with [N~{\sc
  ii}] contours superposed, Panel (b) shows [N~{\sc ii}] $\lambda$6583
  in grayscale and contours obtained at roughly the same slit offset
  position as the near-IR lines (see Fig.\ 1), and Panel (c) shows
  H$_2$ contours over the [N~{\sc ii}] emission in grayscale.}
\end{figure}

% FIGURE 4 ------------
\begin{figure}
\epsscale{0.4}
\plotone{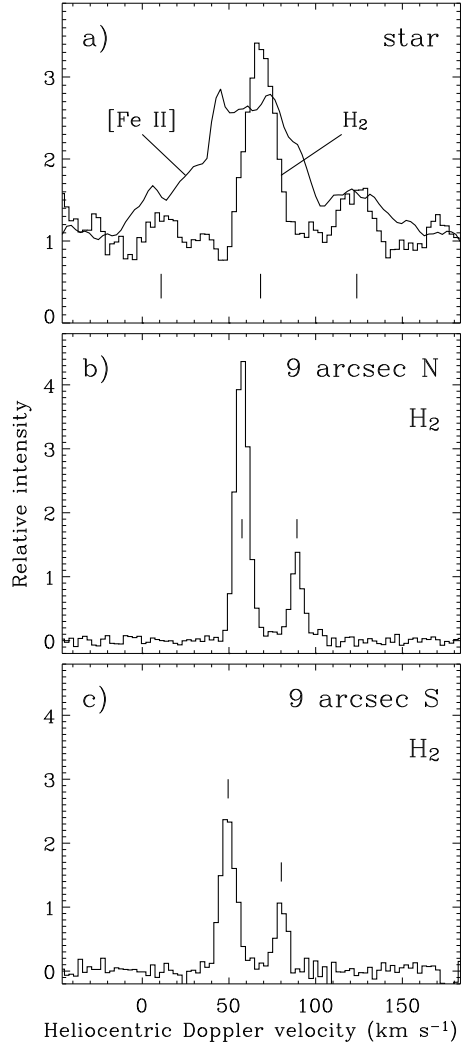}
\caption{Tracings of [Fe~{\sc ii}] $\lambda$16435 and H$_2$
  $\lambda$21218 (histogram) near the central star (a), as well as
  tracings of H$_2$ in the N and S polar lobes (Panels b and c,
  respectively), at positions along the slit between 8$\arcsec$ and
  10$\arcsec$ from the central star.  Tick marks show the measured
  velocities of H$_2$ listed in Table 1.  The relative intensity scale
  is arbitrary.}
\end{figure}

% FIGURE 5 --  ---- SKETCH ----------
\begin{figure}
\epsscale{0.4}
\plotone{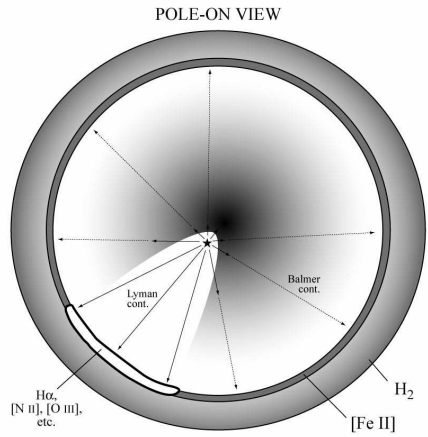}
\caption{A cartoon of one idea for the rotating illumination source in
  M~2-9, explained in \S 4.2.  Solid arrows represent the paths of
  Lyman continuum radiation from the hot star, and dashed arrows are
  paths through the companion's wind along which Lyman continuum is
  extinguished but Balmer continuum penetrates.  The Lyman continuum
  radiation that escapes the wind through the cavity carved by the hot
  star ionizes gas in the lobes, causing the moving patterns of
  H$\alpha$, [N~{\sc ii}], and [O~{\sc iii}] seen in optical images
  over the past several decades. Thus, the inside surfaces of the
  hollow lobes are like a screen illuminated by a rotating
  searchlight.  Balmer continuum that penetrates the wind in other
  directions powers the emission of the [Fe~{\sc ii}] and H$_2$
  shells.  The cavity in the cool star's wind and the corresponding
  escape directions of ionizing radiation will obviously have some
  latitude dependence that is not conveyed in this diagram.}
\end{figure}

\end{document}